\documentclass{IEEEtran4PSCC}
\usepackage{dsfont}
\ifCLASSINFOpdf
   \usepackage[pdftex]{graphicx}

\else

   \usepackage[dvips]{graphicx}

\fi

\usepackage[utf8]{inputenc}

\usepackage[dvipsnames]{xcolor}
\usepackage{colortbl}

\usepackage{footnote}
\usepackage{tablefootnote}

\usepackage{dcolumn}
\newcolumntype{d}[1]{D{.}{.}{#1}}

\usepackage[cmex10]{amsmath}
\usepackage{algpseudocode}
\usepackage[linesnumbered,ruled]{algorithm2e}
\usepackage{cite}
\begin{document}

\title{Electric End-User Consumer Profit Maximization: An Online Approach}

\author{

\IEEEauthorblockN{Arman Alahyari,  David Pozo}
\IEEEauthorblockA{Center for Energy Science and Technology\\ Skolkovo Institute of Science and Technology (\textit{Skoltech})\\
}
}

\maketitle

\begin{abstract}
The fast growth of communication technology within the concept of smart grids can provide data and control signals from/to all consumers in an online fashion. This could foster
more participation for end-user customers. These types of customers do not necessarily have powerful prediction tools or capability of storing a large amount of historical data. Besides, the relevant information is not always known a priori, while decisions need to be made fast within a very limited time. { These limitations and also the novel structure of decision making, which comes from the necessities to make the decision very fast with a limited amount of information, implies a requirement for investigating a novel framework: online decision-making}.

In this study, we propose an online constrained convex optimization framework for operating responsive end-user electrical customers in real-time. Within this online-decision-making framework, algorithms are proposed for two cases: no prediction data is available at the moment of decision-making, and a limited number of forward time periods predictions of uncertain parameters are available.  The simulation results exhibit the capability of the model to achieve considerable profits in an easy-to-implement procedure. Comprehensive numerical test cases are performed for comparison with existent alternative models.   
\end{abstract}



\section{Introduction}


The development of information and communication technology is bringing forward many opportunities toward the smart grid paradigm. In the near future, energy information will be available fast, even in consumers' sites \cite{ij2}. For instance, with 5G technology, data transmission speed can approach to tens of Gbps; 
generating a continuous data stream with enormous amount  of information that would need to be processed on the fly. 
The aforementioned sort of technologies, along with the internet of things and new power electronic devices, will empower customers with vast streams of online data supporting new ways for controlling their consumption; however, at the same time, they will be required to quickly make decisions for exploiting capabilities from their side \cite{k1,4,ij1}. This kind of decision-making belongs to the online-decision making category in which the input is revealed piece by piece and in a serial fashion. In contrast, in the offline decision making all information is required a priori.
\subsection{Literature Review}
Recent avenues of research on different sections of the power system have been focused on real-time or online algorithms for solving power system optimizations and control problems. For instance, the study in \cite{5} addresses real-time power flow and presents a control algorithm to modulate the reactive power output of volt ampere reactive (VAR) resources to minimize total real power delivery at the feeder. This approach does not require the controller to know the network model utilizing a gradient-based extremum to estimate the gradient of the cost function. The study in \cite{6} uses online optimization in optimal frequency where the optimal battery participation in frequency regulation markets is addressed. Indeed, online control policy and an optimal bidding policy based on realistic market settings are presented. The proposed control policy has a threshold structure and achieves near-optimal performance considering a lack of information about future parameters. Voltage regulation in distribution networks is also challenged regarding the penetration of distributed energy resources (DERs) that needs online implementation. In this regard, \cite{7} developed a distributed voltage control (DVC) scheme to achieve the globally optimal settings of reactive power provided by DERs' power electronic interfaces. Further, the DVC design is improved for online implementations that can efficiently adapt to time-varying operating conditions. 

{On the other hand demand response (DR) programs are another section that can potentially enjoy the benefits of online optimization}. 
{A well-known DR model}   {is the real-time demand response} \cite{conejo} {where an end-user customer needs to make decisions about next interval energy consumption within a limited time and with scarce information about future realization of uncertainties. As shown in} \cite{conejo}, {the setting of decision-making process can become  complicated and laying out a comprehensive model and consequently utilizing classic optimization algorithms for each optimization interval (with prediction for all of the remaining time-periods) would become complicated, time-taking and ineffective. Therefore the question rises, since the optimization structure belongs to the online convex optimization (OCO)  framework} \cite{Hazan22,c2,c3}, {could any online approach deal with these shortcomings by applying online algorithms that learns from experience gathered in stages of the optimization. The real-time DR problem can be cast as a constrained optimization problem having two types of constraints, namely: ramp limits on variation of demand and a minimum consumption at the end of the day. Dealing with these types of constraints is not straightforward in OCO context. Therefore several recent studies proposed different approaches each for alleviating the imposed difficulties by these constraints. For instance, OCO with ramp limits are investigated in} \cite{r1,r2,r3} {and OCO with long-term constraints are explained in} \cite{l1,l2,c3}. {Also, more general concepts including dealing with the need for exact constraint satisfaction and  incorporating nonlinear constraints are addressed in} \cite{new1,new2,new3}. {With these recent advances in OCO context, there is a requirement  to investigate the effectiveness of these approaches in the real-time DR application}. 
 Study \cite{10a} presents an online DR from an electric customer point of view. However, the presented online optimization model is a simplified one that does not consider the switching costs, and no prediction capability is incorporated in the proposed online algorithm.

\subsection{Problem Statement and Paper Contributions}
In this study, we  investigate an online DR problem from an electric end-user point of view. We assume that the customer receives a stream of data regarding electricity price. The decisions are taken one by one for every interval. 
{The online DR model considered in our work possesses a structure with long-term and ramp constraints, which complexifies the OCO problem that needs to be solved.} We first introduce the mathematical framework of OCO with the aforementioned constraints in separate subsections. Also, since we consider prediction capability, a proper online algorithm is utilized with a limited look-ahead window of prediction for uncertain parameters. Then, a general framework of an online algorithm for solving the real-time DR problem is introduced. We modify and apply the introduced online algorithm to solve the online DR model for an electric end-user consumer. In order to validate our algorithms for DR, a comprehensive numerical study is given which includes the comparison of different methods, such as: classical OCO with no predictions, OCO including predictions with different length of prediction windows, OCO with incorporated accelerated Nesterov approach, rolling-window robust optimization method and offline optimization with perfect hindsight.

{In summary, the contributions of this study are as follows:}
 \begin{enumerate}   \item {Propose a through online optimization framework and the required algorithms based on the recent developments of online convex optimization studies to deal with the problem of the real-time DR in power system for two cases of with and without incorporating predictions.}
    \item {Analyzing and evaluating the capabilities of the proposed algorithms by presenting a thoroughly comparative set of numerical studies. Benchmarks with popular approaches from literature are also assessed.} 
\end{enumerate}

The remainder of this paper is as follows. Section \ref{optimizationModel} {presents mathematical fundamentals on online decision making needed for understanding the algorithms for online DR introduced in Section} \ref{sec. DRmodel}. Section \ref{numerical} provides numerical studies and section \ref{con} concludes the paper.

\section{{Mathematical Preliminaries on Optimal Online Decision-Making}}\label{optimizationModel}

\subsection{Online Convex Optimization}
\begin{figure}[!h]
\centering
  \includegraphics[width=\linewidth,trim={0cm 0 0 0cm},clip]{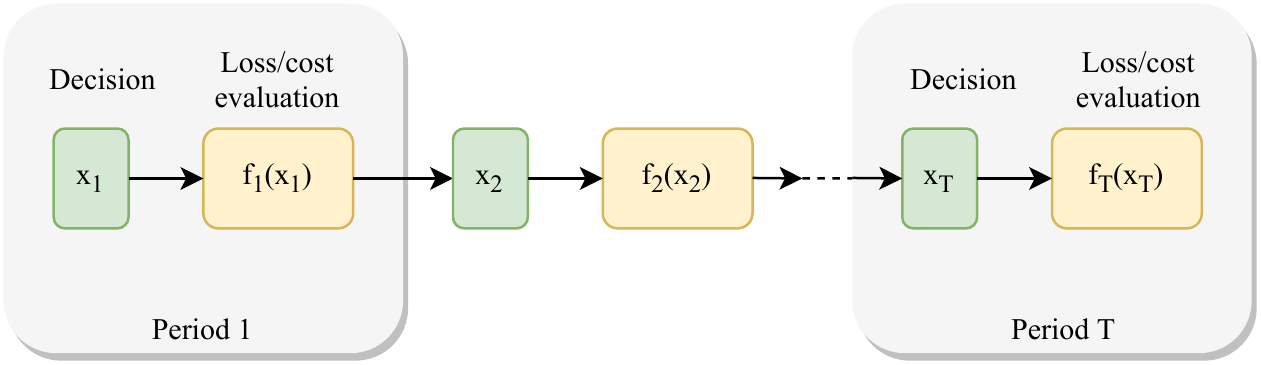}
  \caption{Online optimization decision-making framework.}
  \label{OCO}
\end{figure}

It is important to distinguish online optimization from other similar categories of optimization. {In online optimization, decision-making is performed in a sequence of $T$ (number of optimization rounds) consecutive rounds}. At each step or round, an online decision-maker is supposed to provide the decision for the next round  (for instance, an amount of energy to buy). Then after committing to a specific value, the uncertain parameters are realized. Subsequently, the decision-maker can measure its performance (by a loss function) and maybe update the decision-making process for future rounds. This structure is illustrated in Fig. \ref{OCO}. Indeed, at each stage $t$, the decision-maker determines a decision noted by $x_t$, then the related loss function, $f_t(x_t)$, is realized. According to this loss function, the decision for the next stage is updated. This procedure continues until the last stage of the optimization.

Note that online optimization does not generally mean a high-speed optimization, but it needs to be in a format and sequence that is explained above. Typically, the loss function is uncertain, and future information is scarce or null. The decisions can be taken in a streaming fashion (no need for significant memory usage), and, depending on the optimization needs, fast algorithms can be applied to solve the required optimizations.

\subsubsection{Objective in Online Optimization}

The optimization goal is to minimize the accumulation of costs (loss function) at each step over-optimization period $T$:

\begin{equation}
   C_{\text{online}}= \sum_{t=1}^Tf_t(x_t)
\end{equation}

This overall cost, $C_{\text{online}}$, can be used as a reference to evaluate the performance of online optimization algorithm compared to other existing methods. It is worth to mention that online optimization provides a sub-optimal solution of the offline optimization version, where information for all parameters are known, and a single (offline) problem is solved accordingly.

\subsubsection{Basic Definitions for Online Optimization}
One of the important types of online optimization, which recently has been under attention, is online convex optimization. In OCO the loss function, $f(x)$, is convex with the condition depicted in \eqref{eq.conv1}. However, it can also have properties given in \eqref{eq.oco2} and \eqref{eq.oco3}.
\begin{equation}
\label{eq.conv1}
    f(y) \geq f(x)+ \nabla f(x)(y-x), \quad \forall x,y \in \mathcal{K}
\end{equation} 
\begin{equation}
  f(y) \geq f(x)+ \nabla f(x)(y-x)+\frac{\alpha}{2}\|y-x \|^2, \: \forall x,y \in \mathcal{K}  \quad \label{eq.oco2}  
  \end{equation}
  \begin{equation}
  \| f(x) - f(y) \| \leq G\|x-y\|,\: \forall x,y \in \mathcal{K}  \label{eq.oco3}
  \end{equation}
In \eqref{eq.oco2} the condition regarding the $\alpha$-strongly convexity of $f(x)$ is displayed and \eqref{eq.oco3} dictates that $f(x)$ is Lipschitz continuous with parameter G. 
Note that in OCO, the decisions are chosen from a convex set in Euclidean space denoted by $\mathcal{K}  \subseteq \mathds{R}^d$.

The online gradient descent (OGD) is, per se, the cornerstone algorithm for solving the OCO. It is described as follows:

\begin{algorithm}\label{alg.OGDs}
\SetAlgoLined
\DontPrintSemicolon
  \textbf{Inputs} $x_0 \in \mathcal{K}$, $\eta \in \mathds{R}^+$\;
  \textbf{for} {$t=1$ to $T$} {\textbf{do}}\;
    \quad Compute: $x_{t}$= $x_{t-1} - \eta \nabla f(x_{t-1})$\;
  \quad  Project:  $x_{t}$ =  $\prod_{\mathcal{K}}(x_{t})$\;
  \textbf{end}

 \caption{Online Gradient Descent}
\end{algorithm}
Typical problems that are solved within the OCO concept are not constrained. 
\subsubsection{Constrained OCO}
Many optimization problems belong to the OCO category. However, the real-world applications of these optimization problems introduce specific constraints that need to be considered in the online model. 

These constraints may only affect the decision-making process after a certain number of rounds, known as long-term constraints or are needed to be firmly addressed at each step of the decision-making, such as ramp constraints. Our DR model is an application with both types of the aforementioned constraints, we explain a general structure of OCO with long-term  constraints and the ramp constraint are introduced while presenting the demand response model. The online algorithms could also incorporate prediction of the uncertain parameters. In this regard, we present algorithms that utilize a certain number of look-ahead prediction windows. Then, we propose online DR algorithms to solve the online DR model with and without considering predictions.


\subsubsection{Long-Term Constraints}
As the name already implies, this type of constraint only needs to be fulfilled after several rounds. That means the constraint can be violated before the specific period. By assuming that there is only one long-term constraint, the decision-set can be defined as follows.
 \begin{equation}
    \mathcal{K}=\{x\in \mathds{R}^d: \sum_{t=1}^{T}g_t(x_t)\leq 0\}
\end{equation}

It is assumed that $g_t(x_t)$ is Lipschitz continuous and the decision domain is bounded.

An approach that can be utilized is to move from a constrained optimization to an unconstrained one by introducing a penalty term in the objective function example of which can be $f_t(x_t)+\delta_t g_t(x_t)$. 
In the OCO context, this approach has shown not to be very effective. Thus, other modifications/augmentations of the loss function have been suggested \cite{20l}. For instance,

\begin{equation}\label{eq.long2}
    F_t(x_t)= f_t(x_t)+\delta_t g_t(x_t)-\frac{\gamma\delta_t^2}{2}.
\end{equation}

Here, in equation \eqref{eq.long2}, $\frac{\gamma\delta_t^2}{2}$ is a regularizer preventing large values of $\delta_t$ that in addition to $x_t$ needs to be decided in the online algorithm and the decision updating. Updating can be carried utilizing OGD as shown in \eqref{eq.gd1} and \eqref{eq.gd2}. Note that, there is a minimization with respect to the primal variable $x$ and a maximization with respect to the dual parameter $\delta$.

\begin{IEEEeqnarray}{ll}
   x_t & = x_{t-1}-\eta \nabla_x F_{t-1}(x_{t-1}, \delta_{t-1}) \label{eq.gd1} \\
 \delta_t &= \delta_{t-1}+\mu \nabla_\delta F_{t-1}(x_{t-1}, \delta_{t-1})  \label{eq.gd2}
\end{IEEEeqnarray}

Parameters $\eta$ and $\mu$ are optimization updating pre-fixed stepsizes. Observe that, $\gamma$ is not a stepsize; however, it is fixed and tuned in the optimization to avoid large variations of $\delta_t$.

\subsubsection{Forward-Looking Algorithms}

Clearly, one cannot hope for an online algorithm to
perform optimal if nothing is known about future
cost functions, especially in a case of ramp constrained optimization since the current choice places limits on
feasible choices in the next stage. Thus, it is possible also to consider a situation
where the algorithm has information about both the current
cost function and a limited number of future cost functions.

Indeed, in OCO with the forward-looking algorithm, it is assumed that the algorithm has perfect look-ahead information for $W$ steps. 

With the above assumption, at each stage decision-maker chooses $x_t$ after observation of the cost functions of $W$ forward windows: [$f_t(x_t), ... ,f_{t+W-1}(x_{t+W-1}) $].


\subsubsection{Specialized Algorithms for Solving OCO}
A wide variety of algorithms can be utilized within OCO problems. However, considering the look-ahead features of the aforementioned model, perhaps
the most natural candidates are algorithms
from the control community such as Model Predictive Control
(MPC) and its variations.
Nevertheless, the classic MPC approaches require solving a large optimization problem at each stage, which is usually computationally expensive.
 
Thus, variation of MPC has been utilized in the recent literature in related OCO problems including
forward-looking algorithms: Averaging Fixed Horizon Control (AFHC) \cite{r1, rr2}, Receding Horizon Gradient Descent (RHGD) and Receding Horizon
Accelerated Gradient (RHAG)\cite{rh}. Considering the desired structure in our study, we utilize the latter modifying it to include long-term and ramp constraints and then apply it to the introduced online DR model. 

Both of RHGD and RHAG algorithms are basically adapted from offline gradient-based algorithms: gradient descent and Nesterov’s accelerated gradient \cite{nest} and have similar frameworks. In fact, RHAG only differs in how the gradient is utilized to update the next stage decision.

\textit{RHGD framework} assumes that there exists a forward window of information with the length of $W$.  The algorithm includes two different sections, first to initialize decisions utilizing gradient descent and then updating them $W$ times with gradient descent before reaching the final decision. Each step of updating utilize forward information to help reaching the closer to optimal decisions. In this regard, in RHGD algorithm, to calculate decision $x_t$ at stage $t$, the determination process begins at stage $t-W$, $W$ stages before $t$,  where initial value of decision $x_t$ is calculated, denoted by $x_{t}^{t-W}$. Next, for stages $t-W+1,...,t$, RHGD updates decision $x_t$ denoted by $ x^{t-W+1}_t,...,x^{t}_t$. The final decision will be the output for stage $t$ denoted as $x_t^{t}$ \cite{rh}.


\textit{RHAG framework} uses the same structure while instead of gradient descent updates shown in \eqref{eq.gd1}, Nesterov accelerated gradient is applied in each stage to update decisions as depicted in \eqref{eq.nest11}.

\begin{equation}  \label{eq.nest11}
\begin{split}
  x_t&= y_{t-1}-\eta\nabla_x F_{t-1}(y_{t-1})\\
  y_t&=(1+\xi) x_t-\xi x_{t-1}
\end{split}
\end{equation}
Here, $\xi=\frac{1-\sqrt{\zeta\eta}}{1+\sqrt{\zeta\eta}} $ which value of $\zeta$, like other stepsize parameters is assumed to be previously known.

\section{{End-User Demand Response Modeling}}\label{sec. DRmodel}

\subsection{{Offline Demand Response Model}}

From an end-user electric consumer perspective, there is one exogenous signal that is important enough to shape the energy consumption profile of the day: electricity price. {For DR programs, this signal is usually a real-time price, i.e., provided in the order of a few minutes.} However, consuming energy provides a certain amount of utility to the customer. This value could be constant or vary over the optimization period.
Thus, it is important to consume the right amount of energy to minimize costs \cite{ij3}. In this regard, the following offline DR model \cite{conejo} is defined for the end-user customer.
\begin{IEEEeqnarray}{r'l}
    C^*_{\text{offline}} = \min_{\mathbf{x}}  &  \sum_{t=1}^T   \left( \lambda_t  x_t   - U_t(x_t)   \right)  \qquad \quad     \label{objective}\\
     \text{s.t. } & \sum_{t=1}^T  x_t   \geq E_{T} \label{eq.min}\\
     & x_{t}-x_{t-1} \leq r^{up} \qquad  \forall t= 1,\ldots,T   \label{eq.r1} \\
    & x_{t-1}-x_{t} \leq r^{dn} \qquad  \forall t= 1,\ldots,T  \label{eq.r2}\\
    & \underline{x}\leq x_{t} \leq \overline{x}  \qquad \qquad \forall t= 1,\ldots,T  \label{eq.domain}
\end{IEEEeqnarray}

Here, {equation} \eqref{objective}  {presents the objective of the optimization where} $\lambda_t$ is energy price, $x_t$ is consumed energy by the customer (decision variable) and $U_t(x_t)$ represents the utility provided after consuming energy amount of $x_t$ at time $ t$. {Note that, decision variables are the sequence} $\mathbf{x}=\{x_1, x_2, ..., x_T\}$ of energy procurement.
The consumer needs a minimum energy consumption at the end of the period. Implying that even if there exist many high price periods and it becomes more beneficiary to consume less amount of energy (in term of cost) still, at the end of the day, to provide consumer needs a minimum amount of energy should be utilized as depicted in \eqref{eq.min}.
Due to electric infrastructures limitations, power consumption cannot change arbitrarily at each period meaning there is actually a ramp limit on the power consumption variation at each interval.
Ramps limits links energy consumption between two consecutive periods as shown in \eqref{eq.r1} and \eqref{eq.r2}. 
Also, this energy should stay within its limits that actually specifies the general decision-making domain denoted by \eqref{eq.domain}.


\subsection{Online Demand Response Algorithms}

{
There are some remarks that need to be pointed out regarding the  limitations of the offline DR model} \eqref{objective}--\eqref{eq.domain}.
{In general, the end-user consumer has no information about the upcoming future, meaning that no price of electricity is known a priori. This is especially relevant in horizons of 24 hours (typical horizon in related literature) with real-time price signals of 5 minutes. Even if the end-user customer has statistical information, e.g., a probability distribution or uncertainty support sets, it would be useless or counterproductive in the sense that variability incorporated to the offline (stochastic/robust) models will result in poor out-of-sample performance.
Another weak assumption from offline optimization modeling is to schedule the energy commitments for the day-ahead in intervals of 5 minutes. While this is clearly difficult for many end-user customers to fulfill, it could also represent a missing economic opportunity for very volatile real-time spot prices.}
    %
    %

{Motivated by these limitations, we present the following steps that occur in sequence for the online decision making algorithms.}

\begin{itemize}
    \item[\textbf{S1}] {At each $t$,  \textit{select} energy procured $x_t$ that satisfies $x_t \in \mathcal{B}_t$  using an algorithm $\mathcal{A}$. 
    Below, we present two families of online algorithms with the option of considering predictions or not. }
    \vspace{0.2cm} \\
    {The set $\mathcal{B}_t \subseteq \mathcal{K}$ is defined according to the feasible decision space for $x_t$, i.e., $\mathcal{B}_t=\{x_t\in \mathcal{K}: (x_{t-1}-r^{dn})\leq x_t \leq (x_{t-1}+r^{up})\}$. Note that, for a univariate case and $r^{dn}=r^{up}=r$, the set $\mathcal{B}_t$ translates into a graph with length of $2r$ centered with $x_{t-1}$ known while updating $x_t$.} 
    \vspace{0.2cm} 
    \item[\textbf{S2}] {Observe electricity price $\lambda_t$ and utility function $U_t(\cdot)$. } 
    \vspace{0.2cm} 
    \item[\textbf{S3}] {Compute cost/loss incurred in $t$ by }
   \begin{equation}
        f_t(x_t) = \lambda_t  x_t - U_t(x_t). \nonumber
  \end{equation}
  {After step \textbf{S3}, the process goes back to \textbf{S1} until eventually  the $T$ steps are reached.} 
   \vspace{0.2cm} 
  \item[\textbf{SF}] {In the final step, compute the overall accumulated cost/loss incurred by algorithm $\mathcal{A}$ with } 
  \begin{equation}
    {C_{\text{online}}(\mathcal{A}) = \sum_{t=1}^T f_t(x_t).}
    \label{eq.onlineDR}
    \end{equation}
\end{itemize}
  

 { It is worth to observe that  $C_{\text{online}}(\mathcal{A}) \geq  C^*_{\text{offline}}$ is always satisfied. We would like to find algorithms that exhibit performance close to the offline optimal costs.}  

Next, we present algorithms to deal with the introduced online DR model with two basic assumptions: whether the information about the future is available or not.

\subsubsection{{Online DR with no prediction}}
When there is no available data for the upcoming uncertain parameters, the online algorithm should only rely on the previous data and revealed loss functions and update decisions one by one per stage. The online DR algorithm with no prediction is depicted in Algorithm \ref{alg.a}. The stepsizes $\eta$ and $\mu$, and parameter $\gamma$ are considered as inputs. Since no information is available about the future, the optimization begins with an initialization which determines from what point decisions of the algorithm evolve. With the start of the optimization at each stage $t$, the previously updated decision is submitted. Afterward, energy price would be known and loss function is determined according to the utility of customer. Then, by adding the long-term constraint penalization to the loss function, the decision for the energy consumption of the next stage can be calculated. This energy variation is limited to ramp constraints. Therefore, decisions for energy are projected back to $\mathcal{B}_t$, which is centered by previous value of energy and ramp limits as radius of the decision set at time $t$. Values of $\delta_t$ are updated accordingly to make sure that long-term constraint would hold at the end of the optimization period. Values of $\delta_t$ are selected within $\mathds{R}^+$ domain. Note that algorithms here are only given for one interval. However, the decision-making process follows the same pattern for other optimization periods and is continued until the last round, $T$.

\begin{algorithm}\label{alg.a}
\SetAlgoLined
\DontPrintSemicolon

\textbf{Inputs}  $\eta \in \mathds{R}^+, \mu \in \mathds{R}^+,\gamma \in \mathds{R}^+$\;
   \textbf{Initialize} \;
   \quad Collect $x_{t-1}$ \;
   \quad Build $\mathcal{B}_{t}$\; 
    \textbf{Update} \;
  \quad Compute and project \;
  \quad $x_{t}$ = $\prod_{\mathcal{B}_{t}}(x_{t-1} - \eta \nabla_x  F_{t-1}(x_{t-1}, \delta_{t-1}))$\;
  \quad Compute and project \;
  \quad $\delta_{t}$ = $\prod_{\mathds{R^+}} (\delta_{t-1} + \mu \nabla_\delta
   F_{t-1}(x_{t-1}, \delta_{t-1}))$\;
    \textbf{Output} \;
    \quad Return: $x_{t}$, $\delta_{t}$ 
 \caption{Online DR with No Prediction for $t$}
\end{algorithm}

\subsubsection{{Online DR when predictions are available for $W$ stages}}
These values are then utilized for backward updating of the decision variables. 

In the previous case due to the unavailability of applicable data (i.e., future values of optimization parameters) algorithm could only incorporate one step of gradient descent to finalize the value of the next interval decision. 

In this section, we assume that $W$-long look-ahead window of information is available about the uncertain parameters (electricity prices). Thus, algorithm could exploit this information to reach better results. In doing so,  we incorporate a new cost into the previously defined function of $F$ as given in \eqref{eq.newobj}. In this way, the difference between two consecutive decisions is penalized with utilization of parameter $\rho$. While updating values of energy for each stage algorithm avoids hasty steps; therefore, it remains closer to the domain defined by the ramp constraints.

\begin{equation}
\label{eq.newobj}
\begin{split}
   \mathcal{F}_t(x_t, \delta_t)&=  \lambda_t  x_t   - U_t(x_t)  + \frac{\rho}{2}\|x_t-x_{t-1}\|^2\\
   &+\delta_t (\frac{E_T}{T}-x_t)-\frac{\gamma\delta_t^2}{2}
\end{split}
\end{equation}

Note that, with this new function $x_t$ and $x_{t-1}$ are related. Thus, while minimizing total cost $\mathcal{F}^T=\sum^T_{t=1}\mathcal{F}_{t}$, $\nabla_{x_t}\mathcal{F}^T$ no longer equals to $\nabla_{x_t}\mathcal{F}_t$. In fact, it equals to:

\begin{IEEEeqnarray}{rl}
    \nabla_{x_t}\mathcal{F}^{T} & = h_t(x_{t-1},x_t,x_{t+1},\delta_t)=  \label{eq.L} \\
& \left\{
	\begin{array}{ll}
		\nabla_{x_t}F_t(x_t,\delta_t)+\rho(2x_t-x_{t-1}-x_{t+1})  & \mbox{if } t \leq T \\
		\nabla_{x_t}F_t(x_t,\delta_t)+\rho(x_t-x_{T-1}) & \mbox{if } t = T
	\end{array} \nonumber
\right.
\end{IEEEeqnarray}  


Thus, according to the above mentioned gradient step energy in $t$ is dependent on its value in $t+1$. However, since there is at least one look-ahead window and the initialization step provides values of energy in the look-ahead period, the introduced gradient is computable.

\begin{figure}[!h]
\centering
  \includegraphics[width=\linewidth,trim={0cm 0 0 0cm},clip]{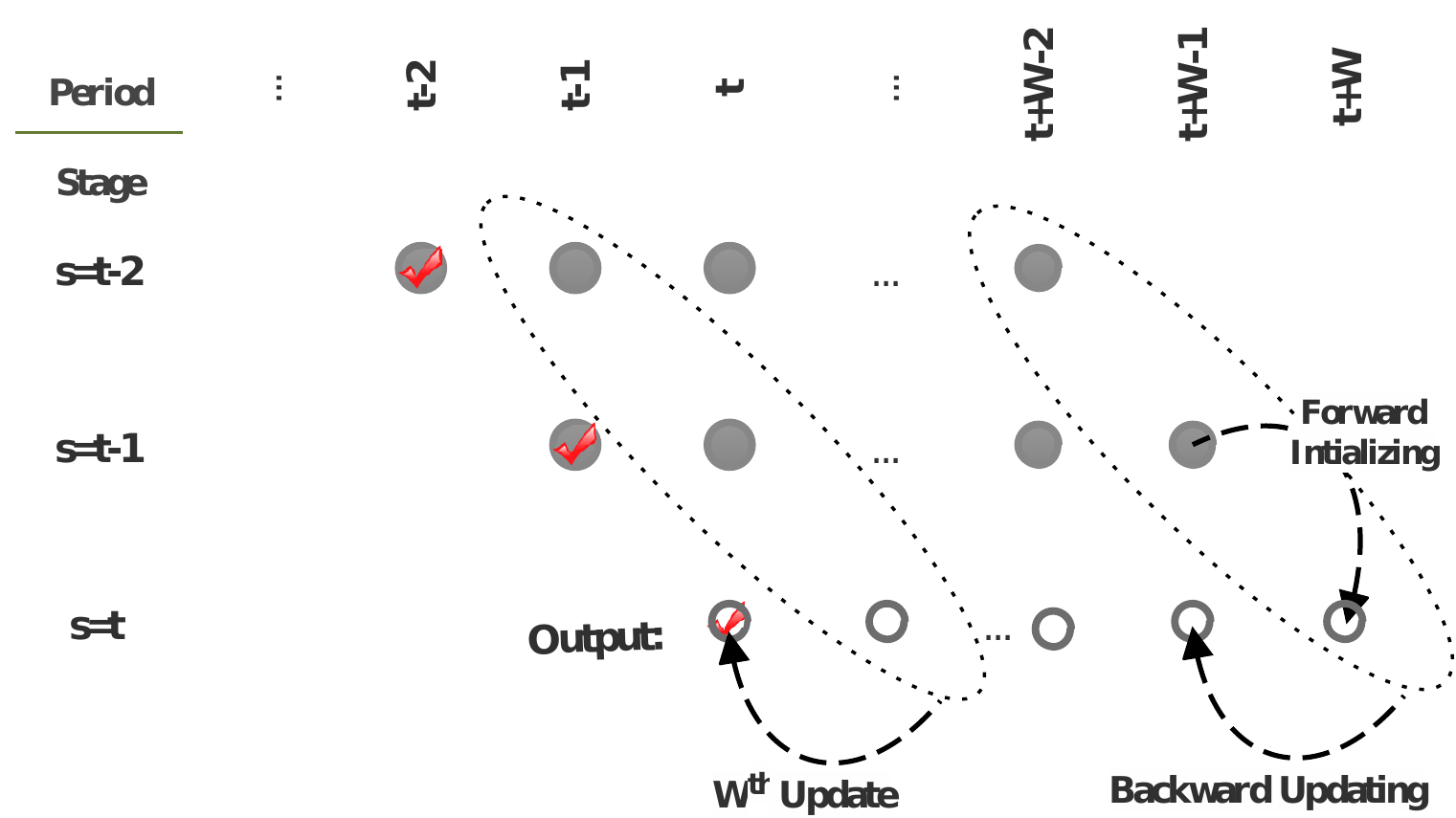}
  \caption{The overall updating procedure  for online RHGD- and RHAG-based algorithms at $t$.}
  \label{alg.update}
\end{figure}

Now, we introduce our online DR algorithm with predictions based on the previously explained RHGD method demonstrated as a pseudo-code in Algorithm \ref{alg.1} for period $t$.  
In the first step, when $t=1$, the necessary inputs are introduced to the optimization including initial values for $x$ and $\delta$, number of look-ahead windows, $W$, and stepsizes: $\eta_1, \eta_2$ and $\mu_1, \mu_2$. Note that the decision-making variable (electric energy consumption) has both super- and subscript, $x_m^s$. Both $s$ and $m$ are related to $t$. Since $W$ forward predictions exist according to the receding horizon algorithm, $W$ gradient steps can be taken in order to reach the final decision. These steps can be carried backward. Consequently, it is needed to initialize the $t+W$ value first and then update backward, step by step till the last one is reached. Considering the dependence of $h_t$ on $x_{t+1}$ the initialization is carried out with $F_t$
 function and only $x_{t+W-1}^{t-1}$ but for backward steps $h_t$ is utilized. In summary, for each interval of optimization, there is one step of initialization using one value of $x$ and $W$ backward updating steps using $h_t$, which is utilizing three values of $x$ from the previous stages and interval of the algorithm. For example, the first backward updating step after the initialization can be written as follows:

\begin{equation}
\begin{split}
    x^{t}_{t+W-1}&=\\
    &(x^{t-1}_{t+W-1} - \eta_2 h_t(x^{t-2}_{t+W-2}, x^{t-1}_{t+W-1}, x^{t}_{t+W},\delta_{t+W-1}^{t-1}))
    \end{split}
\end{equation}
Figure \ref{alg.update} is also given to better clarify the decision-making process. In this figure, dark circles represent the decisions determined already in the previous periods, and white circles are the new updates with the goal of finalizing the output for period $t$. As can be seen, in updating decision at $t$, $s$ is only taking values of $t,t-1,t-2$ and $m$ from $t+W$ to $t$. Although the difference between two consecutive value of energies is penalized, to make sure that ramp limits are not violated, at the last step when the final decision is determined, $x_t^t$, it is projected back to a smaller set constrained by ramp limits.

Next, we introduce online DR with utilization of RHAG in Algorithm \ref{alg.rhag}. As can be seen, it only differs from the previous algorithm when updating the energy values backwardly in term of updating gradient steps. The framework is similar to introduced online  RHGD-based model; therefore, we refrain from adding extra explanation.

\begin{algorithm}\label{alg.1}
\SetAlgoLined
\DontPrintSemicolon
 \textbf{Inputs}  $\rho$, $W$, $\eta_1 \in \mathds{R}^+$, $\eta_2 \in \mathds{R}^+, \mu_1 \in \mathds{R}^+,\mu_2 \in \mathds{R}^+ ,\gamma \in \mathds{R}^+$\;
 
    
  
   \textbf{Initialize} $x^{t}_{t+W}$ and $\delta_{t+W}^{t}$: \;
   \quad Compute and project $x^{t}_{t+W}$= $\prod_\mathcal{K}(x^{t-1}_{t+W-1} - \eta_1 \nabla_x
   F_{t+W-1}(x^{t-1}_{t+W-1},\delta_{t+W-1}^{t-1}))$\;
   \quad Compute and project $\delta_{t+W}^t$= $\prod_\mathcal{\mathds{R^+}}(x^{t-1}_{t+W-1} - \mu_1 \nabla_\delta
   F_{t+W-1}(x^{t-1}_{t+W-1},\delta_{t+W-1}^{t-1}))$\;
    \textbf{Update backwards}: \;
   \quad \textbf{for} $m$=$(t+W-1)$:$-1$: $t$ \textbf{do}\;
  \qquad Compute and project\;  \qquad $x^{t}_m$=$\prod_\mathcal{K}(x^{t-1}_{m} - \eta_2 h_t(x^{t-2}_{m-1}, x^{t-1}_{m}, x^{t}_{m+1}, \delta_m^{t-1})
  )$\;
   \qquad Compute and project \;
   \qquad $\delta^{t}_m$=$\prod_{\mathds{R^+}}(\delta^{t-1}_{m} + \mu_2 \nabla_{\delta_t}
   \mathcal{F}^{T}(x^{t-1}_{m},\delta^{t-1}_{m}))$\;
   \quad \textbf{end}\;
    \textbf{Output:}\;
   \quad Return: $x_{t}$ = $\prod_{\mathcal{B}_{t}}(x^t_{t})$, $\delta_t^t$\;
 \caption{Online RHGD-based  DR Model for $t$}
\end{algorithm}


\begin{algorithm}\label{alg.rhag}
\SetAlgoLined
\DontPrintSemicolon
 \textbf{Inputs}  $\rho$, $W$, $\eta_1 \in \mathds{R}^+$, $\eta_2 \in \mathds{R}^+, \mu_1 \in \mathds{R}^+,\mu_2 \in \mathds{R}^+ ,\gamma \in \mathds{R}^+$,   \enspace $\xi=\frac{1-\sqrt{\zeta\eta_2}}{1+\sqrt{\zeta\eta_2}} $\;
 
    
  
   \textbf{Initialize} $x^{t}_{t+W}$, $y^{t}_{t+W}$ and $\delta_{t+W}^{t}$: \;
   \quad Compute and project $x^{t}_{t+W}$= $\prod_\mathcal{K}(x^{t-1}_{t+W-1} - \eta_1 \nabla_x
   F_{t+W-1}(x^{t-1}_{t+W-1},\delta_{t+W-1}^{t-1}))$\;
   \quad Compute and project $\delta_{t+W}^t$= $\prod_\mathcal{\mathds{R^+}}(x^{t-1}_{t+W-1} - \mu_1 \nabla_\delta
   F_{t+W-1}(x^{t-1}_{t+W-1},\delta_{t+W-1}^{t-1}))$\;
   \quad $y^{t}_{t+W}=x^{t}_{t+W}$\;
    \textbf{Update backwards}: \;
   \quad \textbf{for} $m$=$(t+W-1)$:$-1$: $t$ \textbf{do}\;
  \qquad Compute and project \; \qquad $x^{t}_m$=$\prod_\mathcal{K}(y^{t-1}_{m} - \eta_2 h_t(y^{t-2}_{m-1}, y^{t-1}_{m}, y^{t}_{m+1}, \delta_m^{t-1})
  )$\;
  \qquad $y^{t}_m$=$(1+\xi)x^{t}_m-\xi x^{t-1}_m$\;
   \qquad Compute and project \;
    \qquad$\delta^{t}_m$=$\prod_{\mathds{R^+}}(\delta^{t-1}_{m} + \mu_2 \nabla_{\delta_t}
   \mathcal{F}^{T}(x^{t-1}_{m},\delta^{t-1}_{m}))$\;
   \quad \textbf{end}\;
    \textbf{Output:}\;
   \quad Return: $x_{t}$ = $\prod_{\mathcal{B}_{t}}(x^t_{t})$, $\delta_t^t$\;
 \caption{Online RHAG-based  DR Model for $t$}
\end{algorithm}

\section{Numerical Studies}\label{numerical}
\begin{figure}[!h]
\centering
  \includegraphics[width=\linewidth,trim={0cm 0 0 0cm},clip]{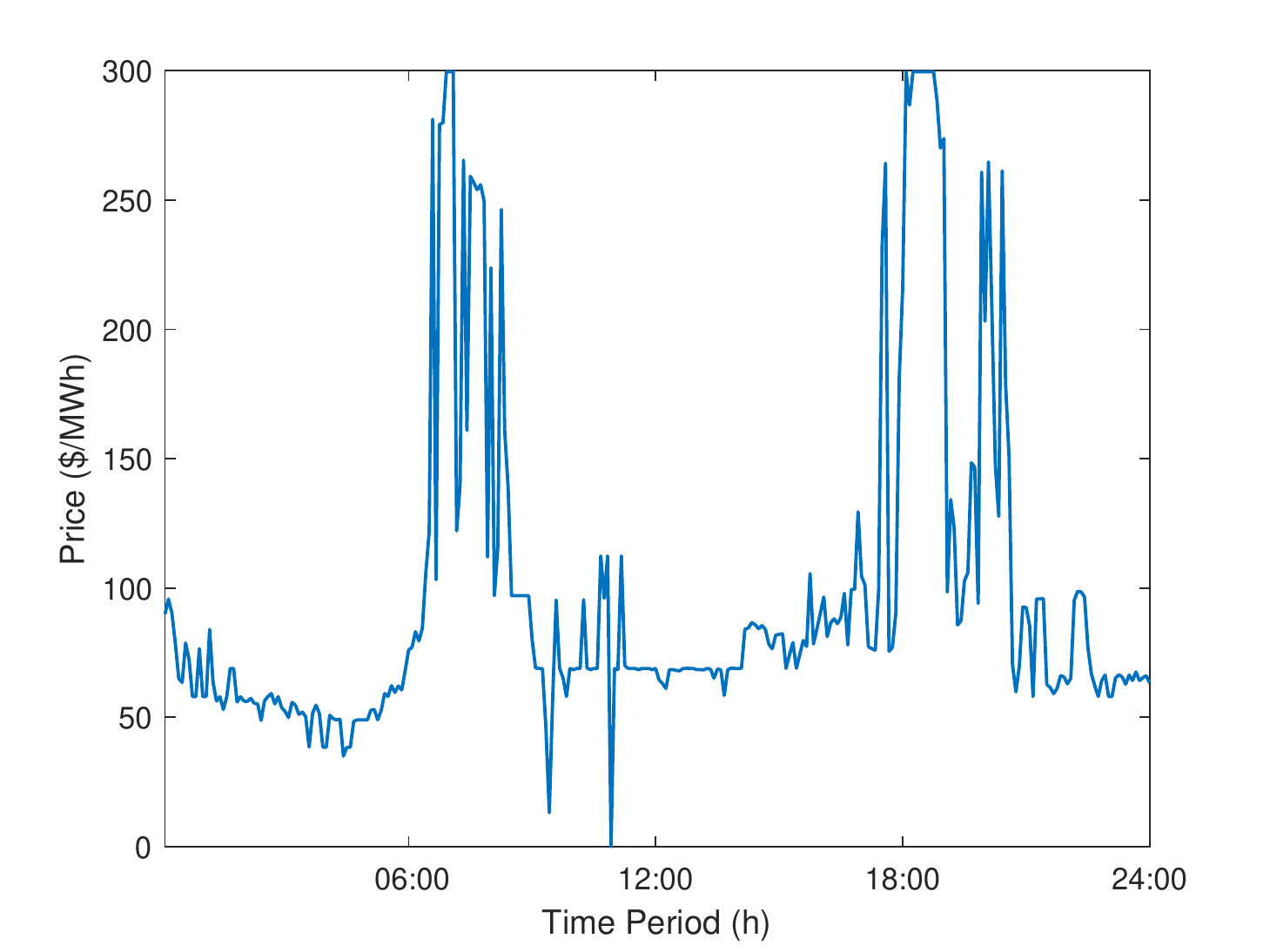}
  \caption{{Five-minute electricity price signal} \cite{price1}.}
  \label{price}
\end{figure}

In this section, we present the numerical studies regarding the online DR model and algorithms introduced in previous sections. We assume an end-user electric consumer that aims to optimize its consumption in a way that the overall cost is minimized throughout the day. The characterization of the consumer load is given in Table \ref{tab. customerData}. From this data, information regarding the decision set, long-term and ramp constraints is available. 
We utilize 5-minute electricity prices from \cite{price1} for an entire day which makes optimization period length
$T$, equal to 288 intervals. This price data is shown in Fig. \ref{price}. We assume consumer receives these prices one by one per interval after committing to an amount of energy consumption. 

{In the remainder of this section, we explore results of five different applicable online-based algorithms namely, (i) online DR without prediction, (ii) online RHGD-based DR with perfect information, (iii) RHAG-based DR with perfect information, (iv) online RHGD-based DR with not exact prediction, and (v) online RHGD-based DR with not exact prediction. In addition, we compare them with two off-line optimization state-of-the-art benchmarks in DR problem, namely (i) a  rolling window robust optimization }\cite{conejo} {(ii) offline optimization with perfect hindsight, i.e., a deterministic case}, the solution of the model \eqref{objective}--\eqref{eq.domain}. Note that, we assume that consumer's utility  acquired by the consumption of energy is calculated by $U_t(x_t)=u x_t$, where $u$ is fixed. In this way, the only uncertain parameter is the upcoming real-time electricity price. We investigate the performance of different algorithms under this assumption. 

\subsection{Online DR without Prediction}

\begin{figure}[!h]
\centering
  \includegraphics[width=\linewidth,trim={0cm 0 0 0cm},clip]{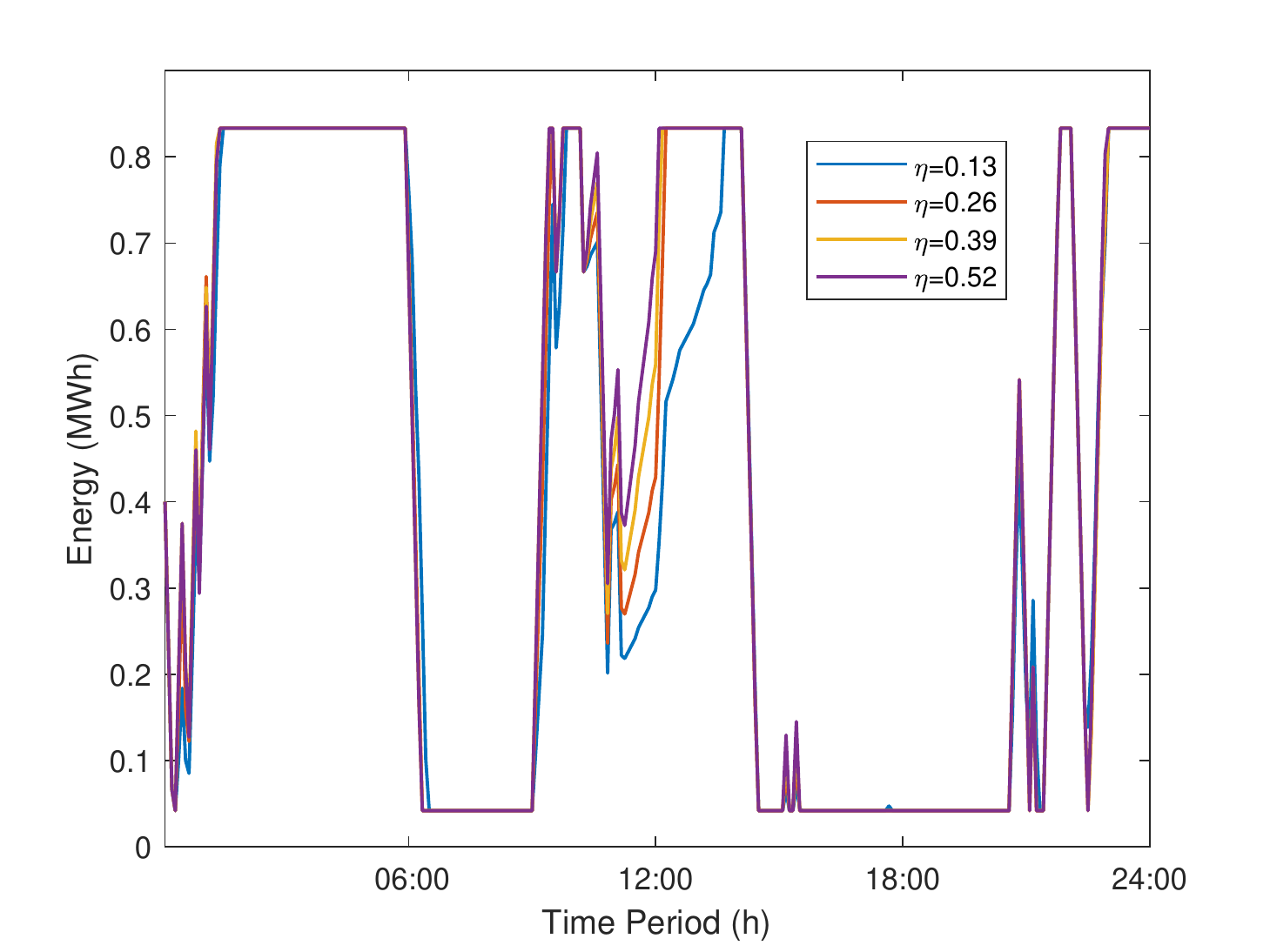}
  \caption{Energy consumption in the optimization period with variation of $\eta$ for online DR with no prediction.}
  \label{noPMS}
\end{figure}

\begin{figure}[!h]
\centering
  \includegraphics[width=\linewidth,trim={0cm 0 0 0cm},clip]{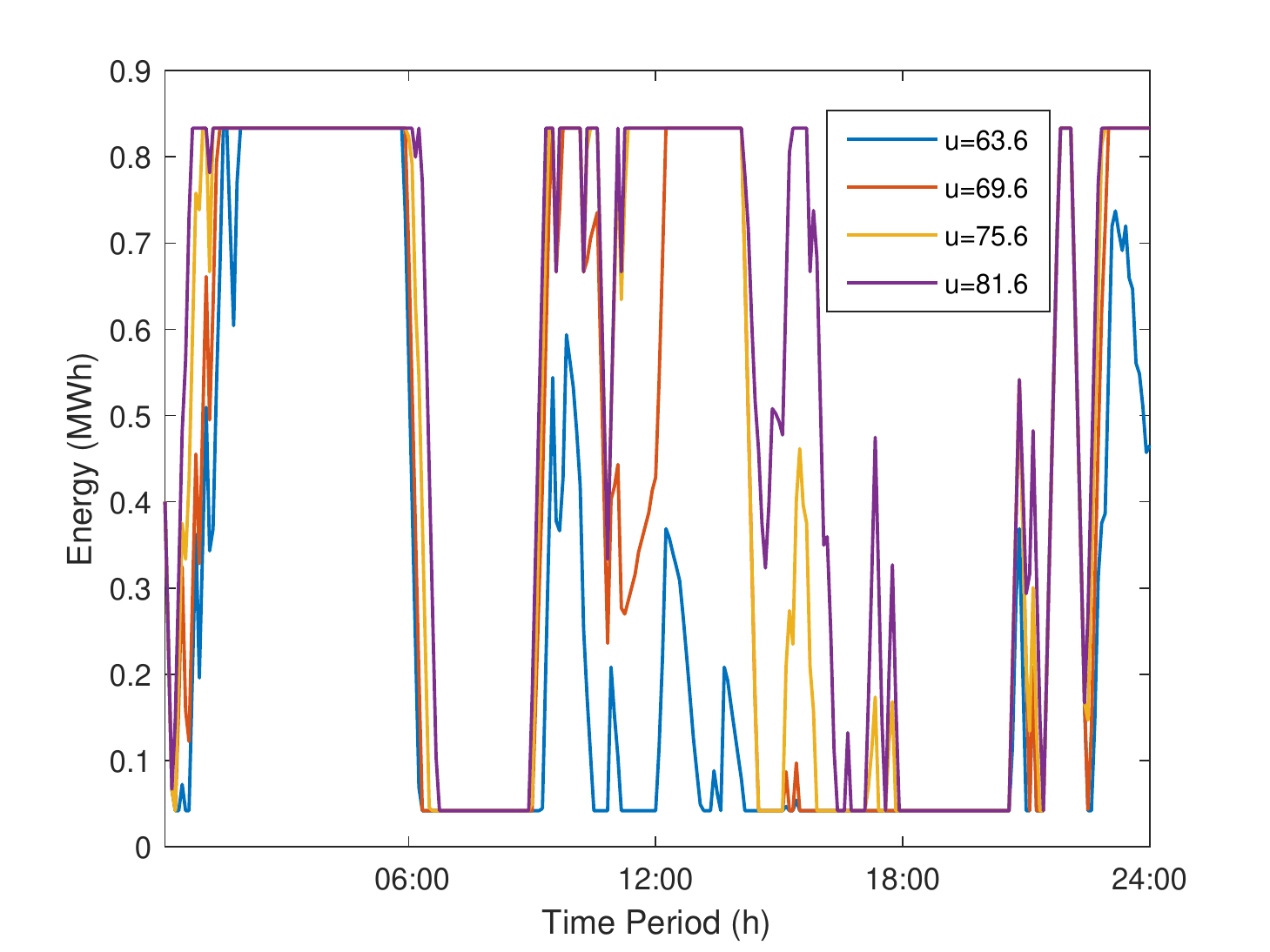}
  \caption{Energy consumption in the optimization period with variation of $u$ for online DR with no prediction and $\eta=0.26$.}
  \label{PMU}
\end{figure}

Algorithm \ref{alg.a} is utilized in this case to manage online DR model without any data about the future upcoming electricity prices. Note that, under different stepsizes, the performance of the Algorithm \ref{alg.a} could vary. To clarify this, we have depicted the performance of the presented algorithm with four different stepsizes: $\eta= \{0.13, 0.26, 0.39, 0.52\}$ in Fig. \ref{noPMS}. As can be seen from this figure, stepsize value affects how much energy can vary from a stage to another. In some periods, the price has fast fluctuations; therefore, bigger changes are needed for energy consumption to adapt to the variations in energy prices. This is more visible at intervals near 12:00.  When stepsize is at the lowest amount, it takes more time for the algorithm to go back to the top position. {However, higher values of stepsizes do not necessarily mean better total profits (total utility minus the total electric energy purchasing cost for the consumer).} For example, the introduced online model will reach  profit of $\{35.43, 38.6, 39.34, 39.16\}$ $\$$ for each stepsize, respectively. It is also of interest to see how the online model would react to different values of utility. In this regard, the energy consumption variation during the optimization period is depicted in Fig. \ref{PMU} considering different values of $u$. It is clear with lower values of consumer utility, the gap between price and utility decreases and it will be more profitable to consume a lower amount of energy which is also perceivable from this figure. As the value of utility increases, it becomes more beneficial to consume more energy. In fact, the gradient has higher positive values resulting in more consumption in favorable intervals. The total profits for these cases are $\{-4.64, 38.6, 103.81, 167.12\}$ $\$$ for lowest to highest value of consumer utility, respectively.

\begin{table}[!h]
\centering
\caption{Consumer electric load characterization data}
 \begin{tabular}{l| l} 
 \hline
 Data  & Quantity  \\ [0.5ex] 
 \hline
 Maximum energy consumption at each interval & 0.834 MWh   \\
 Minimum energy consumption at each interval & 0.041 MWh      \\ 
 Minimum daily consumption & 60 MWh      \\
 Ramping up limit & 2 MW/h     \\
 Ramping down limit & 2 MW/h      \\
 Customer utility & 69.6 \$/MWh      \\
 \hline
 \end{tabular}
 \label{tab. customerData}
\end{table}

\begin{figure}[!h]
\centering
  \includegraphics[width=\linewidth,trim={0cm 0 0 0cm},clip]{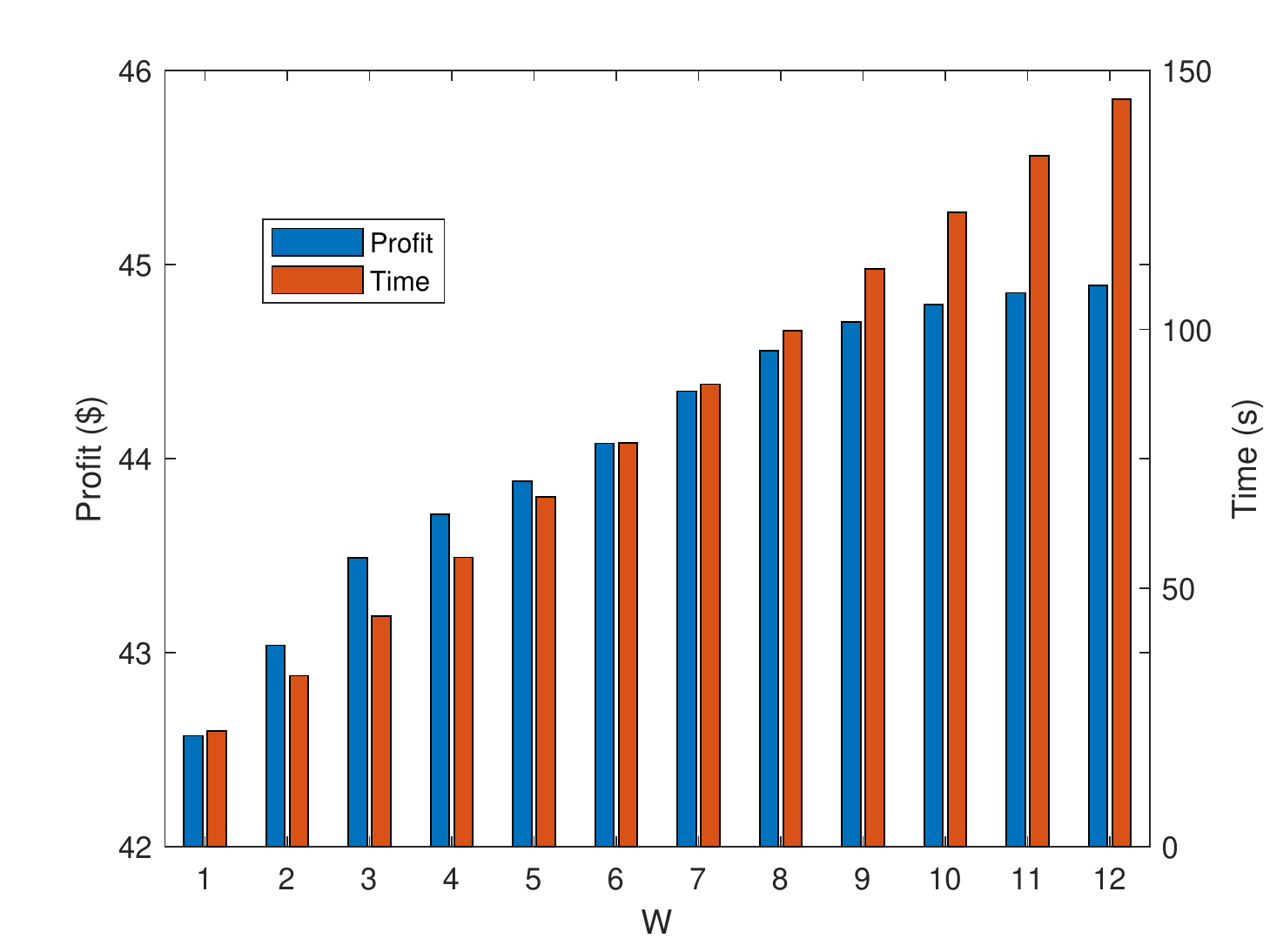}
  \caption{Profit at the end of the optimization and the total optimization time for different look-ahead windows in online RHGD-based DR approach.}
  \label{RHGD}
\end{figure}

\begin{figure}[!h]
\centering
  \includegraphics[width=\linewidth,trim={0.25cm 0cm 0cm 0.5 cm},clip]{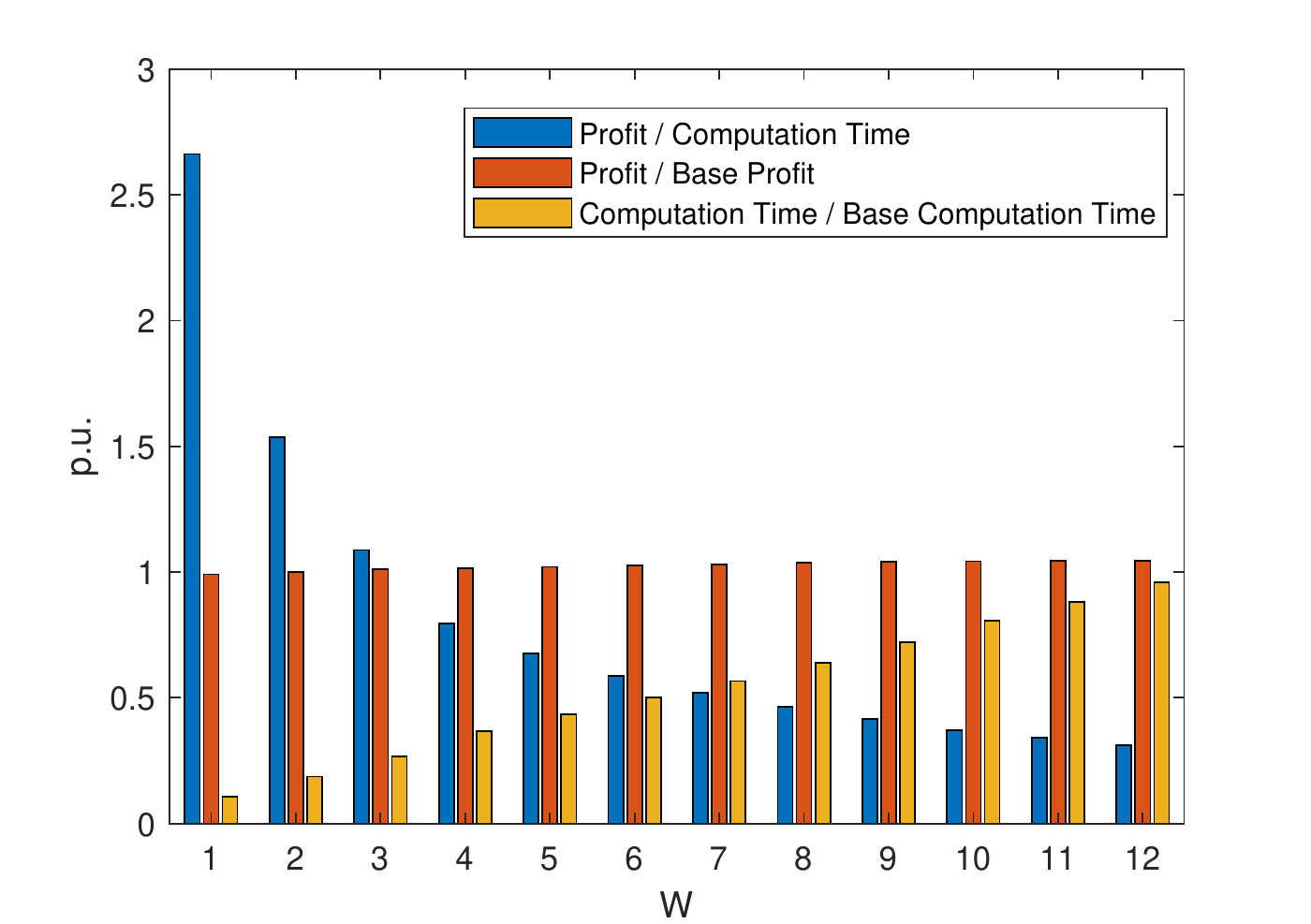}
  \caption{{Ratios level for different length of predictions on profit and time consumption}.}
  \label{RHGDab}
\end{figure}

\subsection{Online DR with Prediction}

\begin{figure}[!h]
\centering
  \includegraphics[width=\linewidth,trim={0cm 0 0 0cm},clip]{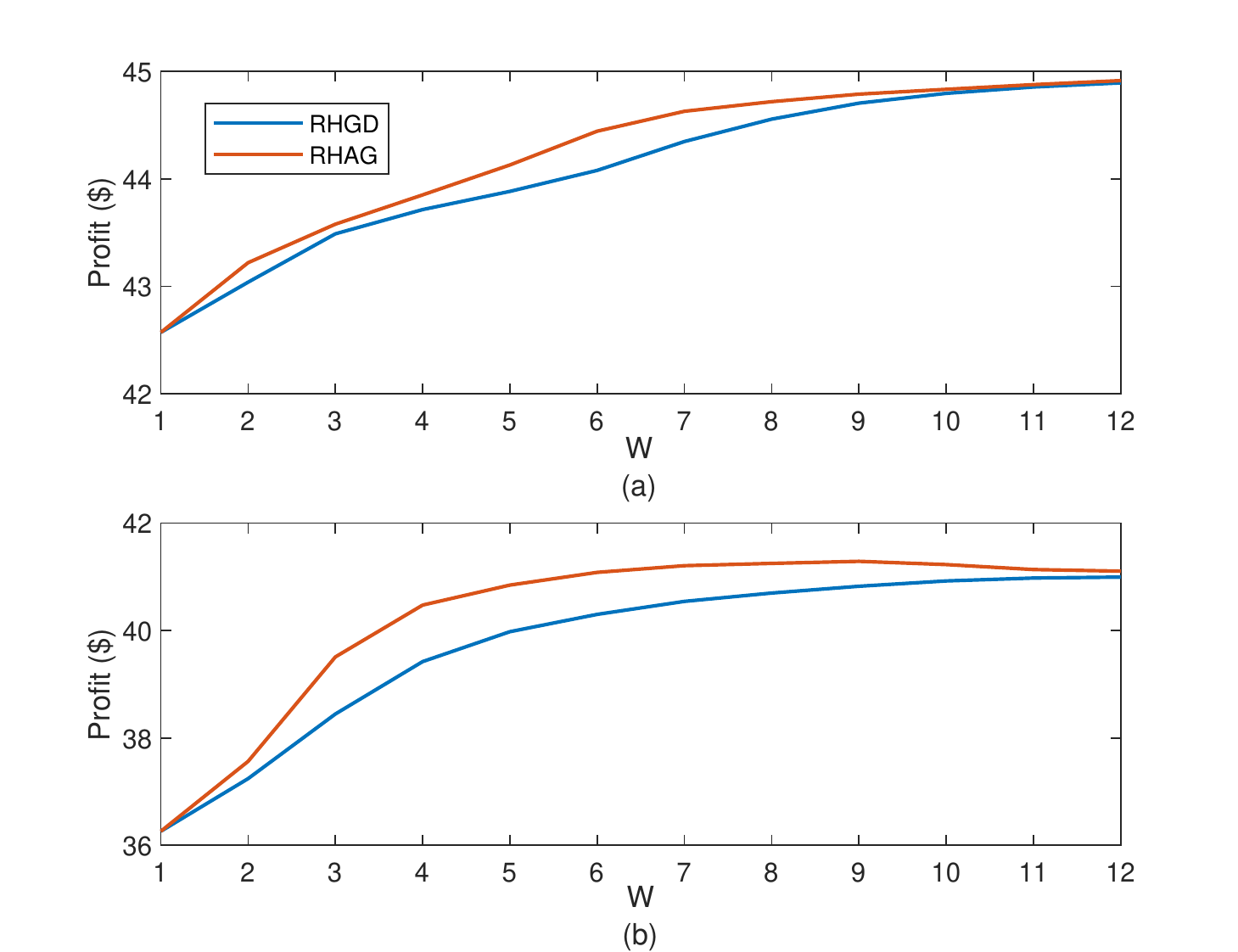}
  \caption{Profits from online RHGD- and RHAG-based DR with limited number of look-ahead information about future prices (a) with prefect information (b) with predictions. }
  \label{GDAG}
\end{figure}

In this part, we analyze the performance of Algorithm \ref{alg.1} and \ref{alg.rhag} where we first assume exact predictions with zero error and then investigate the performance of the introduced online DR models applying not exact predictions. 
In this regard, Algorithm \ref{alg.1} is utilized with different number of perfect look-ahead windows. The total profit for each case (with a specific number of look-ahead windows) are depicted in Fig. \ref{RHGD} along with total time consumed to finish the whole round of the optimization. As can be seen from this figure, even one forward window with perfect information can help noticeably to improve the final result of the model compared to a situation where no information is available. More number of windows of perfect information helps in improving the total acquired profit at the end of the optimization. However, increasing number of look-ahead windows comes with a price which is the fact that by increasing $W$ required time for completing the optimization also increases considerably. For instance, the completion time difference between $W=3$ and $W=12$ is more than 100 seconds. This number would grow exponentially for more complicated optimization problems. Thus, depending on the optimization structure proper value for prediction window should be selected.{ In addition, for the sake of  better comparability in }Fig. \ref{RHGDab}  {three different curves comparing profits and computation times are depicted. In this figure, the base profit and base computation time are assumed to be 43 \$ and 150 s, respectively.}  Next, we have investigated the results of RHAG-based algorithm in Fig. \ref{GDAG} (a). It is clear, RHAG performs better to reach higher values of profit specially with lower number of look-ahead windows. As this number grows,  more iterations are possible and therefore RHGD could reach similar results compared to RHAG. In part (b) of Fig. \ref{GDAG}, an ARIMA model \cite{ar27} is utilized to predict limited number of forward upcoming electricity prices and based on these predictions acquired profits are depicted for both RHGD- and RHAG-based DR models. As can be perceived, the profit values are lower than the case with perfect information. This is due to the fact that gradient of loss function determines direction and size of each step toward updating energy consumption at each stage which itself is directly related to the price value. Thus, imperfect price prediction results in less efficient gradient steps; however the obtained results are comparable to perfect information case. Note that, same as previous case, with lower number of prediction windows RHAG-based approach performs better in term of reaching higher values of profit.

\subsection{Comparison with the DR Offline Optimization Approaches}
\begin{figure}[!h]
\centering
  \includegraphics[width=\linewidth,trim={0cm 0 0 0cm},clip]{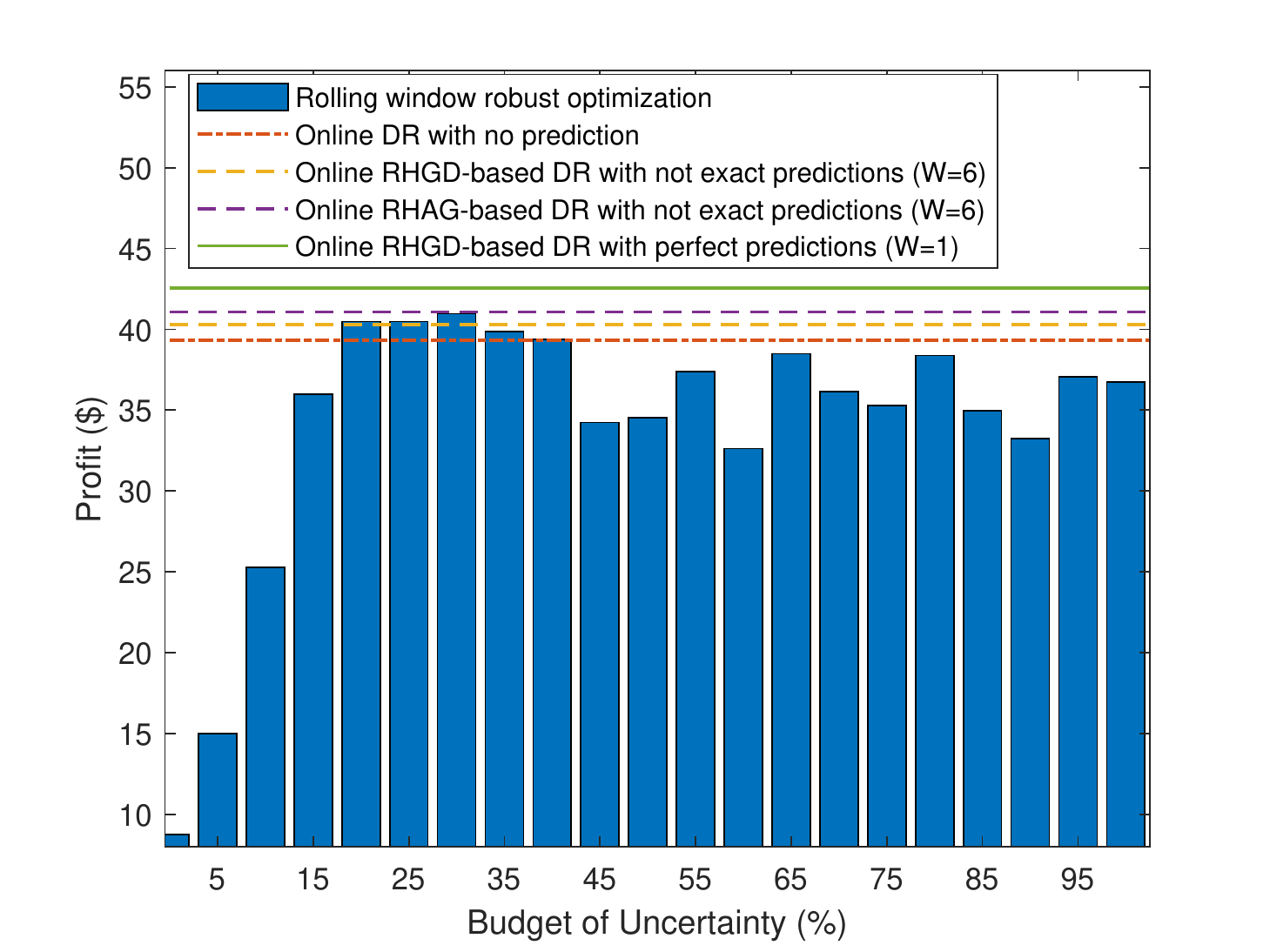}
  \caption{Profits for rolling window robust optimization with different values of uncertainty budget and the proposed online approaches.}
  \label{cono}
\end{figure}

In this section, we compare previous algorithms results with a similar approach utilized in study \cite{conejo} known as rolling-window robust DR model. In this approach, at each stage, a robust optimization problem is solved for the remaining forward periods of the optimization. Thus, it has a similar structure to the introduced DR model. However, it incorporates predictions for all of the remaining optimization stages. Similar to the introduced online DR algorithms, results depend on a tunable parameter mostly known as the budget of uncertainty. The achieved profit while running the DR with this approach considering different values for the budget of uncertainty are demonstrated in Fig. \ref{cono}. The maximum profit of $40.97$ \$ is achieved when uncertainty budget is equal to 0.3. The time needed to finish one round of optimization is 141 seconds. If we compare the result to the one obtained with online DR with no prediction, we see that the acquired profits are pretty close though a little higher for rolling-window robust model. However, in online DR with no prediction the time needed to complete the optimization is only 19 seconds which is 7 time less than the robust approach. When predictions are incorporated into the online DR model  higher profits can be achieved while only limited information about future prices is needed. Note that, with perfect information in hindsight the offline model which is a linear programming model can be carried out resulting in profit value of $53.21$ \$. Meaning that if perfect price information is available for 288 stages the optimization should  achieve this result which is maximum possible for any other methods as well. We showed with only 12 stage of perfect information considerable value of $44.89$ \$ can be achieved in a fast and easy to implement procedure. This also demonstrates the value of good prediction. For example with one window of perfect information the acquired result was higher than any other simulated models with not exact predictions. 

Finally, we have carried out the same procedure for 100 days with different electricity prices. The \textit{average performance} of different algorithms as a percentage of perfect offline optimization are depicted in Table \ref{tab.rf}. Results validate the performance of the proposed online DR approaches. {Our online algorithms with any type of input prediction are more effective (up to 3.5\% more) than the actual state-of-the-art offline robust optimization strategy} \cite{conejo}. {Even when no prediction is used, efficiency reached by our online DR algorithm is close to the offline robust optimization approach. It is encouraging for practitioners, where online algorithms presented here have a much lower computational cost and do not require any off-the-shelf optimization solver.}

\begin{table}[ht!]
\begin{center}
\caption{The average performance of different algorithms as a percentage of perfect offline optimization}
 \begin{tabular}{l| c} \label{tab.rf}
 Algorithm/Approach  & Average  (\%) \\ [0.5ex] 
 \hline
Online DR with no prediction & 69.21   \\
Online RHGD-based DR with not exact prediction (W=6) & \cellcolor{gray!25} 71.91   \\
Online RHAG-based DR with not exact prediction (W=6) & \cellcolor{gray!25} 73.20   \\
Online RHGD-based DR with perfect prediction (W=1) & \cellcolor{gray!25} 75.11  \\
Rolling window robust optimization$^a$, \cite{conejo} & 71.59   \\
Deterministic offline optimization, \eqref{objective}--\eqref{eq.domain} & 100    \\
 \hline
 \end{tabular}
\end{center}
 \footnotesize{{$^a$ Parameter of the budget of uncertainty was selected for the best average performance over the 100 days of simulation}}.
\end{table}

\section{Conclusion}\label{con}
In this paper, we proposed an online demand response model of an electric end-user customer, in which the customer adapts its load with the 5-minute streamed price data in order to reach higher profits while addressing the operational constraints as minimum energy consumption in a day, and ramp constraints on energy variation amounts. 
We investigated two models under the assumption of whether the prediction of future price data was available or not. We then presented online DR models and related algorithms to address the aforementioned problems. The simulation results demonstrated that by utilization of predictions, higher profits could be achieved; however, the computational time also grows accordingly. Comparing the acquired results with those of offline method with perfect hindsight and rolling-window robust optimization showed that the proposed online DR models could achieve considerable profit within a limited time and with very low computational complexity. 
\bibliographystyle{IEEEtran}
\bibliography{myrefs.bib}

\end{document}